\documentclass[preprint,12pt,3p]{elsarticle}

\usepackage{graphicx}
\usepackage{tabularx}
\usepackage{hyperref}
\usepackage{cleveref}
\usepackage{xcolor}

\usepackage{amssymb}
\usepackage{tabu}
\usepackage{longtable}
\usepackage{float}
\usepackage{subfig}
\usepackage{siunitx}
\usepackage{gensymb}
 \usepackage{amsthm}

\journal{Acta Materialia}

\begin{document}

\begin{frontmatter}

\title{Accelerated design of Fe-based soft magnetic materials using machine learning and stochastic optimization}

\author[label1]{Yuhao Wang\fnref{label9}}

\author[label2]{Yefan Tian\fnref{label9}}

\author[label1]{Tanner Kirk}

\author[label3]{Omar Laris}

\author[label2,label4]{Joseph H. Ross, Jr.}

\author[label5]{Ronald D. Noebe}

\author[label5]{Vladimir Keylin}

\author[label1,label4]{Raymundo Arr\'{o}yave\corref{cor1}}

\cortext[cor1]{Corresponding author.}
\address[label1]{Department of Mechanical Engineering, Texas A\&M University, College Station, TX 77843, USA}
\address[label2]{Department of Physics and Astronomy, Texas A\&M University, College Station, TX 77843, USA}
\address[label3]{Department of Materials Science and Engineering, Massachusetts Institute of Technology, Cambridge, MA 02139, USA}
\address[label4]{Department of Materials Science and Engineering, Texas A\&M University, College Station, TX 77843, USA}
\address[label5]{Materials and Structures Division, NASA Glenn Research Center, Cleveland, OH 44135, USA}
\fntext[label9]{Authors contributed equally to this work.}

\begin{abstract}
Machine learning was utilized to efficiently boost the development of soft magnetic materials. The design process includes building a database composed of published experimental results, applying machine learning methods on the database, identifying the trends of magnetic properties in soft magnetic materials, and accelerating the design of next-generation soft magnetic nanocrystalline materials through the use of numerical optimization. Machine learning regression models were trained to predict magnetic saturation ($B_S$), coercivity ($H_C$) and magnetostriction ($\lambda$), with a stochastic optimization framework being used to further optimize the corresponding magnetic properties. To verify the feasibility of the machine learning model, several optimized soft magnetic materials -- specified in terms of compositions and thermomechanical treatments -- have been predicted and then prepared and tested, showing good agreement between predictions and experiments, proving the reliability of the designed model. Two rounds of optimization-testing iterations were conducted to search for better properties. 
\end{abstract}


\end{frontmatter}


\section{Introduction}

\subsection{Motivation}

The pursuit of increased efficiency in energy conversion and transformation requires a new generation of energy materials. Soft magnetic materials are capable of rapidly switching their magnetic polarization under relatively small magnetic fields. They typically have small intrinsic coercivity and are used primarily to enhance or channel the flux produced by an electric current. These alloys are used in a large number of electromagnetic distribution, conversion, and generation devices, such as transformers, converters, inductors, motors, generators, and even sensors. 

In the current materials science community, the accelerated discovery and design of new energy materials has gained considerable attention in light of the many societal and environmental challenges we currently face. Soft magnetic materials are crucial as they are essential elements of electro-magnetic energy transformation technologies. For example, the power transformer is a critical component of the solar energy conversion system, whose performance is ultimately limited by the magnetic properties of the materials used to build the cores. In 1988, Yoshizawa \emph{et al.} presented a new nanocrystalline soft magnetic material referred to as FINEMET, which exhibits extraordinary soft magnetic performance \cite{yoshizawa1988new}. This alloy was prepared by partially  crystallizing an amorphous Fe-Si-B alloy with minor addition of Cu and Nb. This unusual combination of chemistry and processing conditions led to an ultrafine grain structure in an amorphous matrix resulting in excellent soft magnetic properties. The resulting soft magnetic properties of FINEMET type alloys, relevant to electromagnetic energy conversion devices, are a unique combination of low energy losses, low magnetostriction, and high magnetic saturation, up to 1.3 T. This was achieved through an ultrafine composite microstructure of cubic-DO$_3$ structured Fe-Si grains with grain sizes of 10-15 nm in a continuous amorphous matrix, providing a new path for designing next-generation soft magnetic materials.

\subsection{FINEMET-type soft magnetic materials}

The target material system in this work is FINEMET-type soft magnetic nanocrystalline alloys whose properties are categorized into two groups, intrinsic and extrinsic properties. Intrinsic properties include magnetic saturation ($B_S$), magnetocrystalline anisotropy ($K_1$), magnetostriction ($\lambda$), and Curie temperature ($T_C$). $K_1$ and $\lambda$ indirectly influence the hysteretic behavior ($B$-$H$ loop) for each type of core material  by influencing coercivity and core losses of the material.  Extrinsic properties include permeability ($\mu$), susceptibility ($\chi$), coercivity ($H_C$), remanence ($M_r$), and core losses ($P_{cv}$). These are influenced not only by the microstructure, but also the geometry of materials, the different forms of anisotropy, and the effect of switching frequency of the applied fields\cite{willard2013nanocrystalline}.

Among these soft magnetic properties, most can be obtained from its unique hysteresis loop - known as the $B$-$H$ curve, shown in Fig.~\ref{intro}(a),  where B is the flux density generated by an electromagnetic coil of the given material as a function of applied magnetic field strength, H. From this curve the following terms can be defined: 
\begin{figure}
    \centering
    \includegraphics[width=\textwidth]{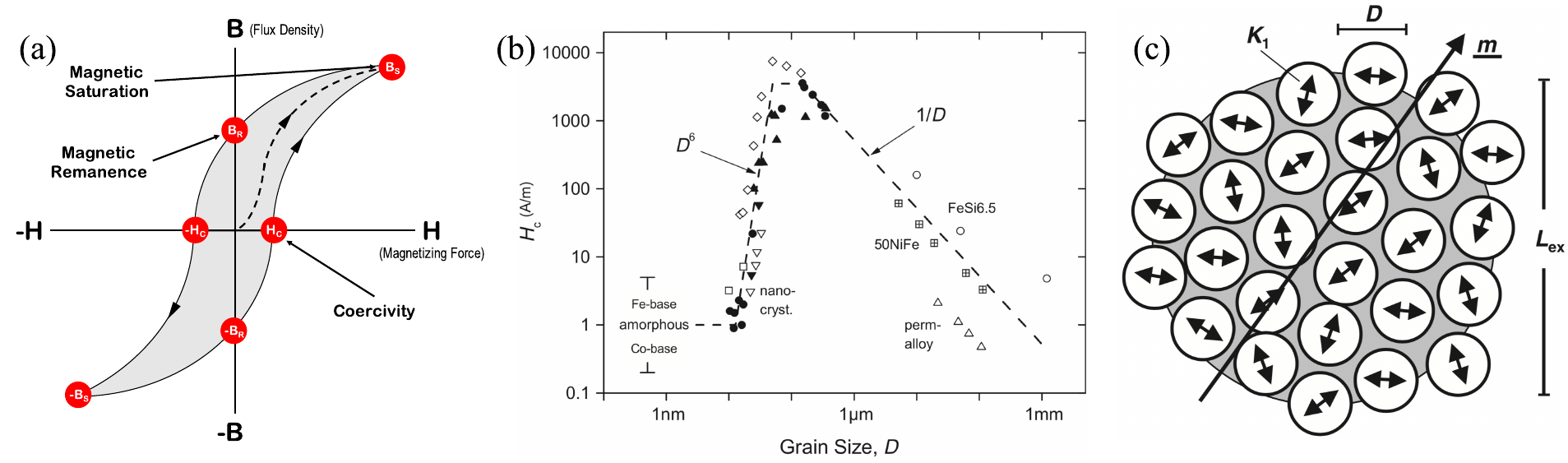}
    \caption{(a) $B$-$H$ loop. (b) Coercivity $H_C$ vs. grain size $D$ for various soft magnetic metallic alloys. Reprinted from Ref.~\cite{herzer2013modern}. (c) Schematic representation of the random anisotropy model for grains embedded in an ideally soft ferromagnetic matrix. The double arrows indicate the random fluctuating anisotropy axis, the hatched area represents the ferromagnetic correlation volume determined by the exchange length $L_{ex}$. Reprinted from Ref.~\cite{herzer2013modern}.}
    \label{intro}
\end{figure}
(a) Coercivity ($H_C$) is the intensity of the applied magnetic field required to reduce the residual flux density to zero after the magnetization of the sample has been driven to saturation. Thus, coercivity measures the resistance of a ferromagnetic material to be demagnetized. (b) Magnetic saturation ($B_S$) is the limit to which the flux density can be generated by the core as the domains in the material become fully aligned. It can be determined directly from the hysteresis loop at high fields. Large values of flux density are desirable since most applications need a device that is light in mass and/or small in volume. (c) Permeability ($\mu$): $\mu=B/H=1+\chi$, is the parameter that describes the flux density, $B$, produced by a given applied field, $H$. Permeability can vary over many orders of magnitude and should be optimized for a given application.  For example, EMI filters usually require large values to produce substantial changes in magnetic flux density in small fields. For other applications, such as filter inductors, permeability does not necessarily need to be high but needs to be constant so that the core does not saturate readily. (d) Core loss is one of the most essential properties of the material as it is a direct measure of the heat generated by the magnetic material in A/C applications. It is the area swept out by the hysteresis loop, which should be minimized to provide a high energy efficiency for the core. Contributions to the core loss include hysteretic sources from local and uniform anisotropies and eddy currents at high frequencies.

Maximizing $B_S$ and minimizing $H_C$ are the most important design objectives for most applications requiring soft magnetic materials and therefore were the design goals in this study. Since $\mu$ heavily depends on the application and can range over several orders of magnitude, even for a fixed composition, depending on secondary processing conditions, it was therefore not a parameter that was optimized or considered in this study. Based on the nature of  FINEMET-type nanocrystalline alloys, several constraints have been incorporated in this study. The magnetic transition metal element has been set to Fe in this work and other elements, such as Co and Ni are excluded,  because at relatively small additions they will tend to decrease B$_S$.  The composition of Fe ranged from 60\%-90\%. The percentages of the remaining  elements in total varied from 10\%-40\%. Although the early transition metal element is Nb in current commercial FINEMET alloys, other elements, such as Zr, Hf, Ta, Mo or even combinations of different early transition metal elements were considered. In commercial alloys,  metalloids B and Si are added to promote glass formation in the precursor and we also allowed for P. The noble metal elements are selected from Cu, Ag, or Au serving as nucleating agents for the ferromagnetic nanocrystalline phase.

The random anisotropy model \cite{herzer1989grain} provides a concise and explicit picture for understanding the soft magnetic properties of nanocrystalline ferromagnetic materials, such as FINEMET. As illustrated in Fig.~\ref{intro}, the microstructure is characterized by a random distribution of structural units or grains in a ferromagnetic matrix with an  effective magnetic anisotropy with a scale $D$. For a finite number ($N$) of grains within the ferromagnetic correlation volume ($V = L_{ex}^3$), the corresponding average anisotropy constant $\langle K_1 \rangle$ is given by
\begin{equation}
    \langle K_1 \rangle \approx \frac{K_1}{\sqrt{N}}=K_1(\frac{D}{L_{ex}})^{3/2},
\end{equation}
which is determined by the statistical fluctuations from averaging over the grains. If there are no other anisotropies, the coercivity $H_C$ and the magnetic saturation $B_S$ are directly related to the average anisotropy constant $\langle K \rangle$ by
\begin{equation}
    H_C=p_c\frac{\langle K \rangle}{B_S},
\end{equation}
where $p_c$ is a dimensionless pre-factor. These relations were initially derived for coherent magnetization rotation in conventional fine particle systems. In the regime $D<L_{ex}$, however, they also apply for domain wall displacements. Accordingly, coercivity was shown to vary with grain size as $H_C∝\propto D^6$ for very fine grained materials (Fig.~\ref{intro}(b)). For large grain sizes, it shows the typical $1/D$-dependence and thus great soft magnetic properties require large grain sizes \cite{pfeifer1980soft}. However, for small grain sizes, $H_C$ shows an extraordinary $D^6$-dependence behavior, which provides another path to realize excellent soft magnetic performance, through the generation of nanocrystalline microstructures \cite{manaf1993new,kneller1991exchange}.

\subsection{FINEMET materials design}

\begin{figure}
    \centering
    \includegraphics[width=\textwidth]{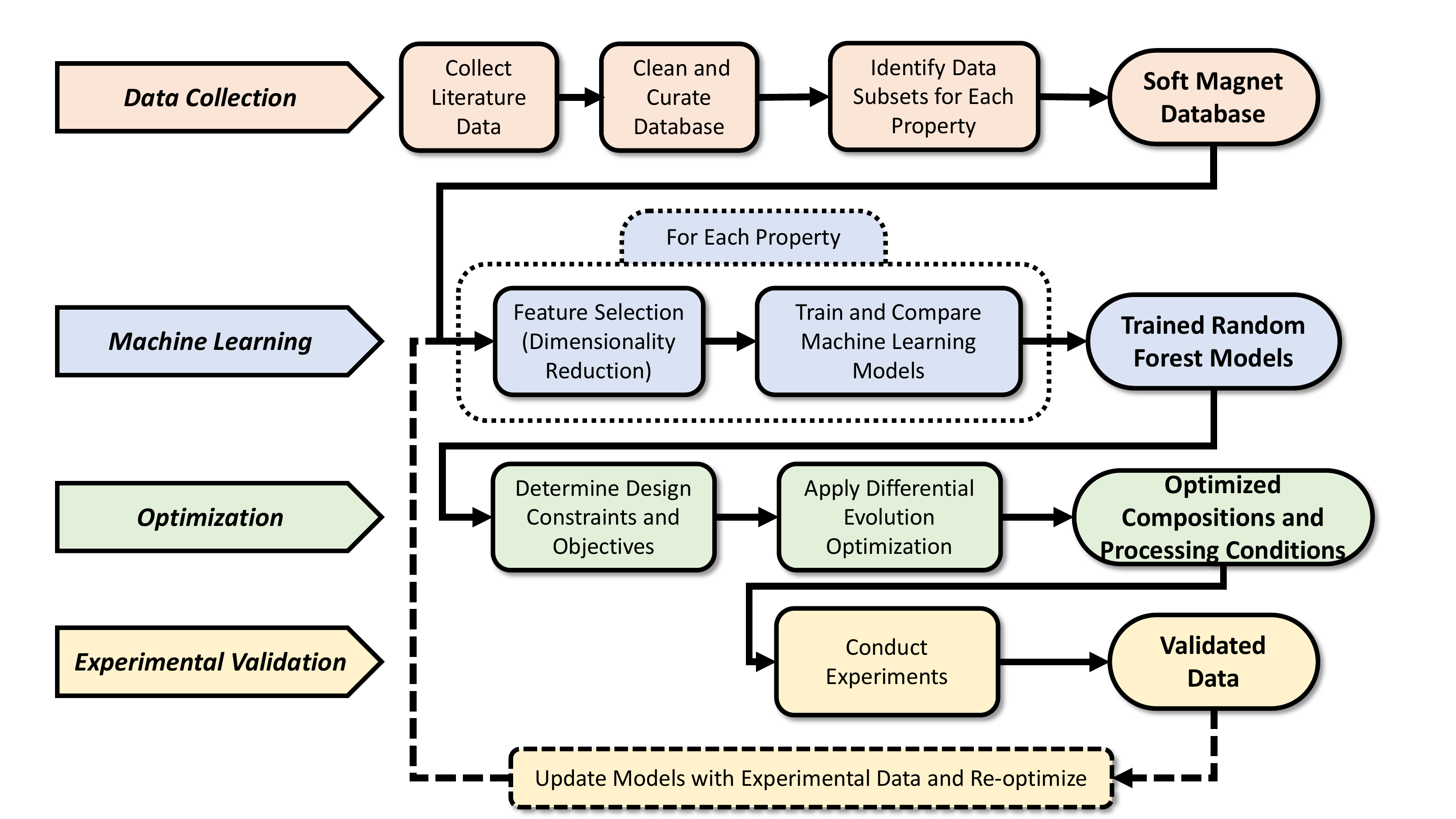}
    \caption{Diagram of machine learning and experimental design framework.}
    \label{framework}
\end{figure}

Machine learning as one of the most popular and efficient statistical techniques has enormous potential to bring the discovery and design of soft magnetic alloys to the next level. Recently, Rajesh \textit{et al}. \cite{jha2018combined} presented a combined CALPHAD and machine learning approach to predict nanocrystalline size and volume fraction from inputs such as composition and heat treatment conditions, demonstrating that machine learning is a promising tool for modeling soft magnetic alloys. Although our work is in a similar material space, we have been focusing on predicting magnetic properties from experimental data mined from the literature. In this work, machine learning was used as the primary technique to expand the boundaries of Fe-based FINEMET-type material space and design the next-generation of soft magnetic materials. As a  first step, we built an experimental database compiled from 76 journal articles published starting in 1988. The database was carefully curated to only include data entries in the nanocrystalline regime. Feature engineering was applied so that the original inputs have been transformed into more general atomic properties.
We built separate models to predict each magnetic property and the number of predictive features were reduced to between 5 and 20 for each property. Various machine learning algorithms were applied to the selected data sets, and the prediction evaluated using 20-fold cross-validation based on the coefficient of determination ($R^2$). Stochastic optimization framework was used to guide the design of new and better soft magnetic nanocrystalline materials. The overall framework is shown in Fig.~\ref{framework}.

\section{Data set overview}

To start the soft magnetic materials design process, the first step was to build a database by collecting available properties from the relevant literature. There are hundreds of accessible publications in the literature base, which describe different types of soft magnetic nanocrystalline alloys in single or a small range of alloy compositions given one or several types of thermal treatments or other types of secondary processing, with some relevant measure of various properties. Due to the limitation of not being able to access the original data directly, it was challenging to collect and organize all the data from the literature to interpret the relations between processing, compositions, structures, and properties. Nevertheless, it is a valuable task to create, maintain and preserve such a data set to share in the materials science community for collaborative research purposes. The sources of our data set are academic literature and patent information, from published experimental work. All publications assembled in the database are listed in Table~\ref{database_literature}. 

\begin{table}
\centering
\begin{tabular}{c c c c} 
 \hline\hline
 References & Year & References & Year \\
 \hline
 Yoshizawa \textit{et al.} \cite{yoshizawa1988new} & 1988 & Saad \textit{et al.} \cite{saad2002crystallization} & 2002 \\ 
 Kataoka \textit{et al.} \cite{kataoka1989soft} & 1989 & Skorvanek \textit{et al.} \cite{skorvanek2002influence} & 2002 \\ 
 Herzer \textit{et al.} \cite{herzer1990grain} & 1990 & Mitrovic \textit{et al.} \cite{mitrovic2002microstructure} & 2002 \\ 
 Suzuki \textit{et al.} \cite{suzuki1990high} & 1990 & Marin \textit{et al.} \cite{marin2002influence} & 2002 \\ 
 Yoshizawa \textit{et al.} \cite{yoshizawa1991magnetic} & 1991 & Zorkovska \textit{et al.} \cite{zorkovska2002role} & 2002 \\ 
 Suzuki \textit{et al.} \cite{suzuki1991soft} & 1991 & Sulictanu \textit{et al.} \cite{sulictanu2002nanostructure} & 2002 \\ 
 Fujii \textit{et al.} \cite{fujii1991magnetic} & 1991 & Cremaschi \textit{et al.} \cite{cremaschi2002evolution} & 2002 \\ 
 Makino \textit{et al.} \cite{makino1991low} & 1991 & Chau \textit{et al.} \cite{chau2003influence} & 2003 \\ 
 Lim \textit{et al.} \cite{lim1993effects} & 1993 & Ponpandian \textit{et al.} \cite{ponpandian2003low} & 2003 \\ 
 Tomida \textit{et al.} \cite{tomida1994crystallization} & 1994 & Kwapulinski \textit{et al.} \cite{kwapulinski2003optimization} & 2003 \\ 
 Makino \textit{et al.} \cite{makino1994magnetic} & 1994 & Crisan \textit{et al.} \cite{crisan2003nanocrystallization} & 2003 \\ 
 Kim \textit{et al.} \cite{kim1995magnetic} & 1995 & Sovak \textit{et al.} \cite{sovak2004influence} & 2004 \\ 
 Inoue \textit{et al.} \cite{inoue1995soft} & 1995 & Cremaschi \textit{et al.} \cite{cremaschi2004magnetic} & 2004 \\ 
 Vlasak \textit{et al.} \cite{vlasak1997influence} & 1997 & Ohnuma \textit{et al.} \cite{ohnuma2005origin} & 2005 \\ 
 Lovas \textit{et al.} \cite{lovas1998survey} & 1998 & Chau \textit{et al.} \cite{chau2006effect} & 2006 \\ 
 Grossinger \textit{et al.} \cite{grossinger1999temperature} & 1999 & Ohta \textit{et al.} \cite{ohta2007new} & 2007 \\ 
 Yoshizawa \textit{et al.} \cite{yoshizawa1999magnetic} & 1999 & Lu \textit{et al.} \cite{lu2008structure} & 2008 \\ 
 Kopcewicz \textit{et al.} \cite{kopcewicz1999mossbauer} & 1999 & Pavlik\textit{et al.} \cite{pavlik2008structure} & 2008 \\ 
 Frost \textit{et al.} \cite{frost1999evolution} & 1999 & Makino \textit{et al.} \cite{makino2009new} & 2009 \\ 
 Franco \textit{et al.} \cite{franco1999magnetic} & 1999 & Ohnuma \textit{et al.} \cite{ohnuma2010stress} & 2010 \\ 
 Turtelli \textit{et al.} \cite{turtelli2000contribution} & 2000 & Butvin \textit{et al.} \cite{butvin2010effects} & 2010 \\ 
 Xu \textit{et al.} \cite{xu2000structure} & 2000 & Lu \textit{et al.} \cite{lu2010microstructure} & 2010 \\ 
 Todd \textit{et al.} \cite{todd2000magnetic} & 2000 & Makino \textit{et al.} \cite{makino2011low} & 2011 \\ 
 Borrego \textit{et al.} \cite{borrego2000devitrification} & 2000 & Kong \textit{et al.} \cite{kong2011high} & 2011 \\ 
 Kemeny \textit{et al.} \cite{kemeny2000structure} & 2000 & Urata \textit{et al.} \cite{urata2011fe} & 2011 \\ 
 Ilinsky \textit{et al.} \cite{ilinsky2000determination} & 2000 & Makino \textit{et al.} \cite{makino2012nanocrystalline} & 2012 \\ 
 Varga \textit{et al.} \cite{varga2000effective} & 2000 & Sharma \textit{et al.} \cite{sharma2014influence} & 2014 \\ 
 Vlasak \textit{et al.} \cite{vlasak2000magnetostriction} & 2000 & Liu \textit{et al.} \cite{liu2015investigation} & 2015 \\ 
 Zorkovska \textit{et al.} \cite{zorkovska2000structure} & 2000 & Wen \textit{et al.} \cite{wen2015structure} & 2015 \\ 
 Solyom \textit{et al.} \cite{solyom2000study} & 2000 & Xiang \textit{et al.} \cite{xiang2015effect} & 2015 \\ 
 Lovas \textit{et al.} \cite{lovas2000saturation} & 2000 & Sinha \textit{et al.} \cite{sinha2015correlation} & 2015 \\ 
 Kwapulinski \textit{et al.} \cite{kwapulinski2001optimisation} & 2001 & Wan \textit{et al.} \cite{wan2016development} & 2016 \\ 
 Borrego \textit{et al.} \cite{borrego2001nanocrystallite} & 2001 & Dan \textit{et al.} \cite{dan2016effect} & 2016 \\ 
 Franco \textit{et al.} \cite{franco2001mo} & 2001 & Li \textit{et al.} \cite{li2017core} & 2017 \\ 
 Mazaleyrat \textit{et al.} \cite{mazaleyrat2001thermo} & 2001 & Jiang \textit{et al.} \cite{jiang2017study} & 2017 \\ 
 Wu \textit{et al.} \cite{wu2001microstructure} & 2001 & Li \textit{et al.} \cite{li2017soft} & 2017 \\ 
 Borrego \textit{et al.} \cite{borrego2001structural} & 2001 & Jia \textit{et al.} \cite{jia2018role} & 2018 \\ 
 Gorria \textit{et al.} \cite{gorria2001correlation} & 2001 & Cao \textit{et al.} \cite{cao2018local} & 2018 \\ 
 \hline\hline
\end{tabular}
\caption{List of soft magnetic papers from which the experimental data were mined.}
\label{database_literature}
\end{table}

\begin{figure}
    \centering
    \includegraphics[width=\textwidth]{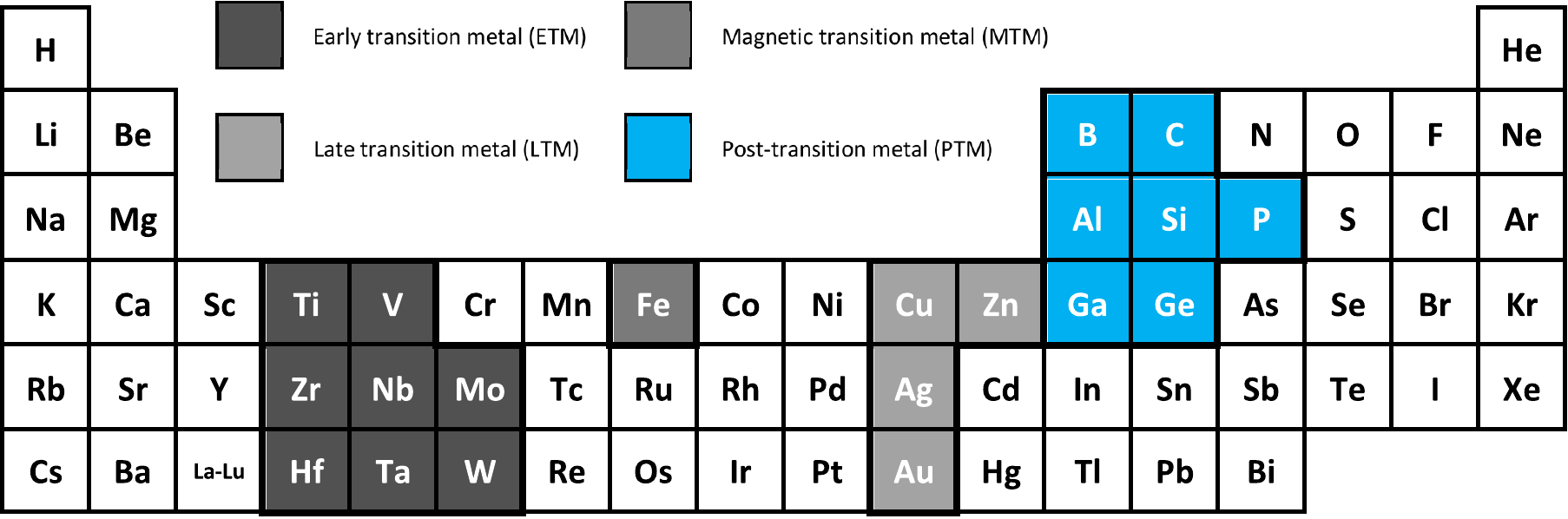}
    \caption{Elemental components of FINEMET-like-soft magnetic alloys  including magnetic transition metal (MTM), early transition metal (ETM),  post-transition metal (PTM), late transition metal (LTM) elements.}
    \label{compositions}
\end{figure}

The essential components of the database are chemical compositions, thermal processing conditions applied to the amorphous precursor ribbon, and soft magnetic properties. Data contained in plots were extracted using \textsc{WebPlotDigitizer} \cite{rohatgi2011webplotdigitizer}. The soft magnetic properties of nanocrystalline materials  are heavily dependent on the chemical composition and the different elements in the composition can be split into four different categories: magnetic transition metal (MTM), early transition metal (ETM),  post-transition metal (PTM) and late transition metal (LTM) as shown in Fig.~\ref{compositions}.
MTM includes Fe, Co, and Ni, but in this first attempt to model behavior, we focused on Fe-based materials, therefore, Co- and Ni-containing alloys were not considered in the database. ETM elements normally act as grain refiners, which help slow down the diffusion process during the thermal crystallization step and help maintain a fine grain size. PTM elements form the glassy phase. In addition, Si serves the critical role of producing the Fe$_3$Si crystalline phase in FINEMET, which has a negative magnetostriction and reduces the overall magnetostriction of the alloy by balancing out the positive magnetostriction in the glassy phase. LTM elements like Cu, Ag, and Au can act as nucleants, which form clusters in the glassy phase, responsible for nucleating a high density of grains.

\begin{figure}
    \centering
    \includegraphics[width=\textwidth]{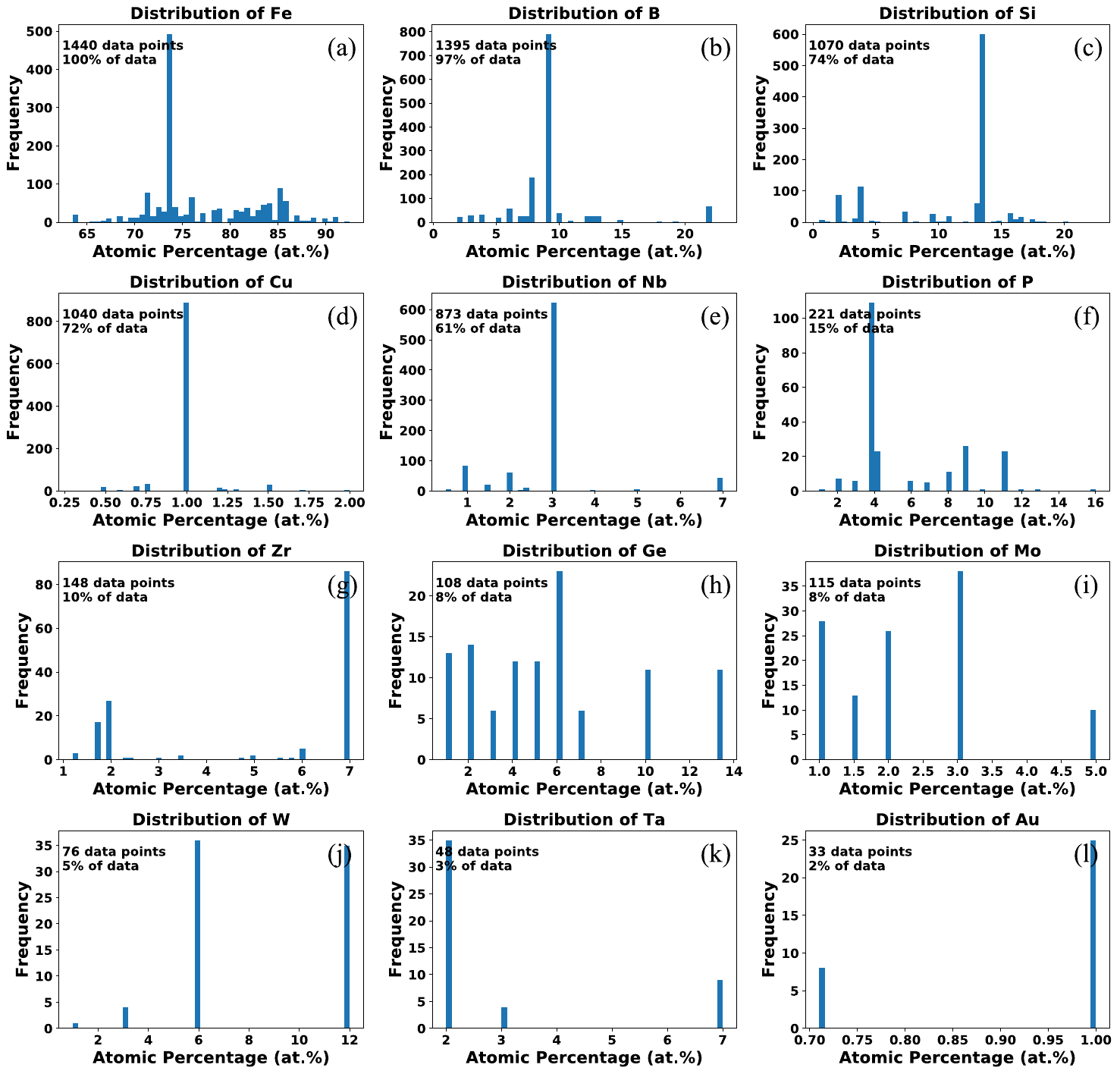}
    \caption{Distribution of various element atomic percentage (at.\%) in the database.}
    \label{distribution_element}
\end{figure}

The distribution of Fe in the FINEMET data set is shown in Fig.~\ref{distribution_element}(a), the dominating composition is 73.5, which is the original atomic percentage of Fe in FINEMET. The distribution of nucleant elements is shown in Figs.~\ref{distribution_element}(b) and \ref{distribution_element}(c). It is shown that only 33 data entries contain Au which is around 2\% of overall entry count. Cu occupies the majority of our studied compositions, and most of the entries contain 1\% of Cu. For ETM elements, the distribution of our study is shown in Figs.~\ref{distribution_element}(h), \ref{distribution_element}(j), \ref{distribution_element}(k), and \ref{distribution_element}(l). The majority of our studied compositions contain Nb as grain refiners and most of them have a Nb fraction around 3\%. The distribution of PTM elements are shown in Figs.~\ref{distribution_element}(d), \ref{distribution_element}(e), \ref{distribution_element}(f), \ref{distribution_element}(g), and \ref{distribution_element}(i). B occurs in most of the entries, and sometimes appears along with P or Ge. The dominant composition for B is 9\%. Si also occurs most of the time because of the need to generate the Fe$_3$Si crystalline phase with a negative magnetostriction and to serve as a potential glass former.

\begin{figure}
    \centering
    \includegraphics[width=\textwidth]{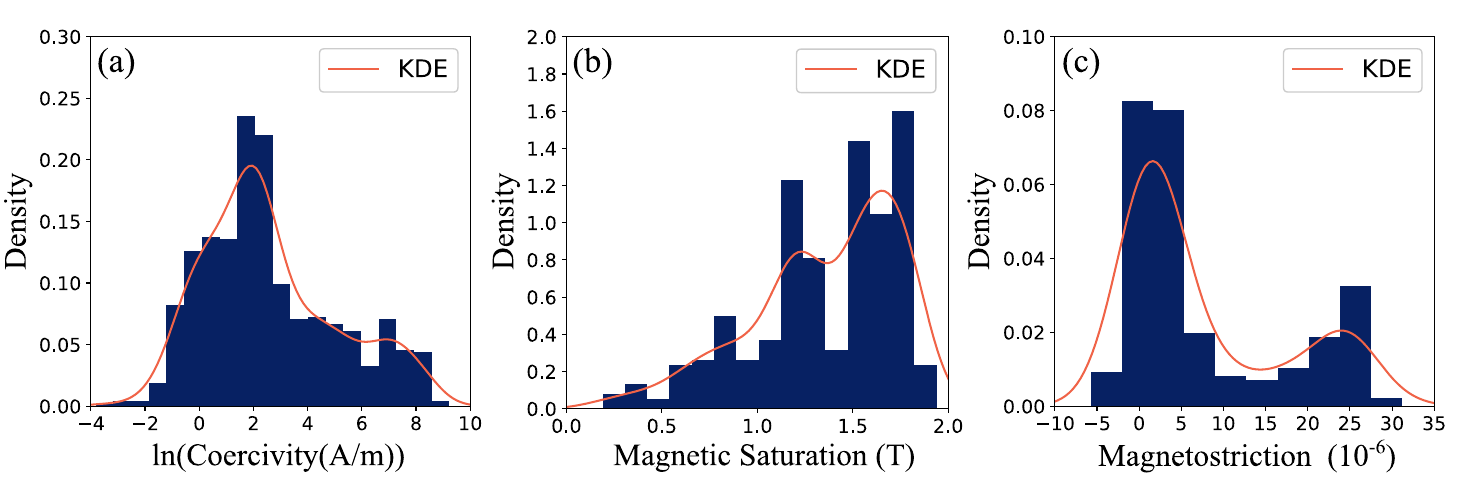}
    \caption{Distribution of selective magnetic properties in the database. Solid lines are kernel density estimations to estimate the probability density function of data entries.}
    \label{distribution_property}
\end{figure}

We also visualize the distribution of selective magnetic properties of interest in Fig.~\ref{distribution_property}.
Because the distribution of coercivity is heavily skewed with a large skewness $\gamma_1=5.1$, we thus applied a log transformation to transform skewed data to approximately conform to normality. The log-transformed data give a much smaller skewness $\gamma_1=0.6$, which can greatly help with the originally observed bias. It is also evident that people tend to report good results, such as high magnetic saturation (larger than 1.5 T) and low magnetostriction (close to 0), rather than ``unattractive" results. In our case, high magnetic saturation data larger than 1.5 T is about 49\% and low magnetostriction data between $-3$ and 3 is 47\%. However, in terms of the machine learning process, the unattractive results can be just as beneficial to the overall process. Due to this reason, we decided not to leave out any ``bad" data. The rest of data (51\% for magnetic saturation and 53\% for magnetostriction) can help with reducing the corresponding biases to a great extent.

In the course of constructing the database, several difficulties, which afflict most materials design problems,  need to be clearly stated. First, we chose FINEMET-type nanocrystalline alloys as the material design space, and there is a vast composition parameter space where few researchers have done measurements. Several clusters of data in the material database represent trendy materials, which are only a small percentage of all candidate material compositions. Second, given the different research habits of various groups and diversity of equipment, some researchers did not clarify everything, such as manner in which magnetic saturation was defined. Finally, not all publications include every soft magnetic property of interest. For example, compared to coercivity, magnetic saturation is more difficult to measure, which leads to the fact that we have many data points for the former and less for magnetic saturation. As a result, considering the primary target is to optimize both coercivity and magnetic saturation, it is helpful to split the database into separate sets, each for a different magnetic property as the objective. All features are listed in Table~\ref{table:features}.
\begin{table}
\centering
\begin{tabularx}{\textwidth}{m{5.5cm}|m{10cm}} 
 \hline\hline
 Chemical Elements & Fe, Si, C, Al, B, P, Ga, Ge, Cu, Ag, Au, Zn, Ti, V, Cr, Zr, Nb, Mo, Hf, Ta, W, Ce, Pr, Gd, U. \\
 \hline
 Experimental measurements & Annealing temperature, Annealing Time, Primary Crystallization Onset, Primary Crystallization Peak, Secondary Crystallization Peak, Longitudinal Annealing field, Transverse Annealing field, Ribbon Thickness, Magnetic Saturation ($B_S$), Coercivity ($H_C$), Permeability ($\mu$), Magnetostriction ($\lambda$), Core Loss, Electrical Resistivity ($\rho$), Curie Temp ($T_C$), Grain Size ($D$). \\ 
 \hline\hline
\end{tabularx}
\caption{Table of all features from the soft magnetic data set.}
\label{table:features}
\end{table}
The features mainly contain two different categories: chemical components and experimental measurements. 

The purpose of the study was to evaluate nanocrystalline FINEMET type alloys.  These alloys are generally produced by partial crystallization of an amorphous precursor by an annealing treatment, which also significantly increases the complexity of the problem. To ensure data consistency, some data points in the data sets were modified by the following procedures:
\begin{enumerate}
    \item Data which were missing an annealing temperature or annealing time, all as-quenched data, and all data processed below room temperature were removed.
    \item Annealing Temperatures were rounded to every 5th degree Celsius.
    \item Annealing Times were rounded to the nearest hour or half hour depending on the magnitude of the value.
    \item Data points out of the nanocrystalline regime. -i.e., grain diameters over 60 nm, were removed.
   	\item Any features, which are unused after data reduction were removed.
\end{enumerate}

\section{Machine learning model}
The next stage in our analysis was to build a series of predictive models using machine learning techniques to relate the magnetic properties with chemical components and processing conditions. The approach consisted of two main steps: feature selection and model selection.
\subsection{Feature selection}
We employed a five step process to identify features that could be removed in order to simplify the model, without incurring a significant loss of information. The procedure includes \cite{feature-selector2018}: (a) Remove the features that have over 50\% missing values. (b) Remove the features containing only one unique value. (c) Identify collinear features and remove them. (d) Remove zero importance features using gradient boosting decision tree algorithm. (e) Remove cumulative low importance features using gradient boosting decision tree algorithm. 

Specifically, step c) utilizes the Pearson correlation coefficient to identify pairs of collinear features. For each pair above the specified threshold (in absolute value), it finds one of the variables to be removed. Steps d) and e), required a supervised learning problem with labels to estimate the importance of features and in this work, a gradient boosting machine implemented in the LightGBM library \cite{ke2017lightgbm} was utilized. A gradient boosting machine works by utilizing multiple "weak learners" (in our case, decision trees) and combining them into a strong learner. Trees are constructed in a greedy manner sequentially and the subsequent predictors can learn from the mistakes of the previous predictors. In a single tree, a decision node splits data into two parts each time based on one of the features, and the relative rank (i.e. depth) of a feature can be used to assess the relative importance of that feature with respect to the target value. Features used at the top level of the tree have a larger contribution to the final prediction based on the input samples. The expected fraction of the samples they contribute to can thus be used as an estimate of the relative importance of the features. In a gradient boosting machine, one can average the importance of the features over a sequence of decision trees to reduce the variance of the estimation in order to use it for feature selection \cite{pedregosa2011scikit}. For step d), features with zero importance were removed. Step e) builds off the feature importance from step d) and removes the lowest important features not needed to reach a specified cumulative total feature importance specified by users. For example, if we input a cumulative total feature importance value 0.99, it can find the lowest important features that are not needed to reach 99\% of the total feature importance and remove them. 

After the feature selection procedure, the remaining features are shown in Table~\ref{table:feature selection}. 
\begin{table}[!htb]
\begin{tabular}{p{.30\textwidth} p{.30\textwidth} p{.30\textwidth}}
 \hline\hline
 Coercivity&Curie Temperature&Grain Size\\
 \hline
Fe&Si&Fe\\
Si&Nb&Si\\
B&Annealing temperature (K)&B\\
P&$T_{c,0}$ (K)&Nb\\
Ge&Ribbon Thickness ($\mu$m)&Annealing temperature (K)\\
Cu&$\Delta T_0$ &Ribbon Thickness ($\mu$m)\\
Au& &\\
Nb&&\\
Mo&&\\
Annealing temperature (K)&&\\
$T_{c,1}$ (K)&&\\
Ribbon Thickness ($\mu$m)&&\\
$\Delta T_1$&&\\
\hline\hline
Magnetic Saturation & Magnetostriction & Permeability\\
\hline
Fe&Fe&Fe\\
Si&Si&Cu\\
B&B&Nb\\
P&Zr&Annealing temperature (K)\\
Zr&Nb&Annealing Time (s)\\
Nb&Ta&Ribbon Thickness ($\mu$m)\\
Annealing temperature (K)&Annealing temperature (K)& \\
Annealing Time (s)&  &  \\
$T_{c,2}$ (K)& & \\
Ribbon Thickness ($\mu$m)& \\
\hline\hline

\end{tabular}
\caption{ Remaining features after feature selection for different materials properties.$T_{c,0}$, $T_{c,1}$, and $T_{c,2}$ represent Primary Crystallization Onset, Primary Crystallization Peak, and Secondary Crystallization peak, respectively. $\Delta T_0$ and $\Delta T_1$ are the difference of $T_{c,0}$ and $T_{c,1}$ with annealing temperature, respectively. }
\label{table:feature selection}
\end{table}
There are 13 features for coercivity; 6 features for Curie temperature; 6 features for grain size; 10 features for magnetic saturation; 7 features for magnetostriction and 6 features for Permeability. Note that initially we tried to add atomic properties to our feature space but later decided to remove them during the feature selection process. It is due to the relatively limited range of element selection we have and the strong correlation between the atomic properties and the chemical compositions. Furthermore, using elemental fraction is more straightforward in the inverse modeling process during optimization, avoiding additional steps in mapping atomic properties back to composition. The following machine learning models were built based on the selected feature sets for each property. 

\subsection{Machine learning algorithms and results}

\begin{figure}
    \centering
    \includegraphics[width=300pt]{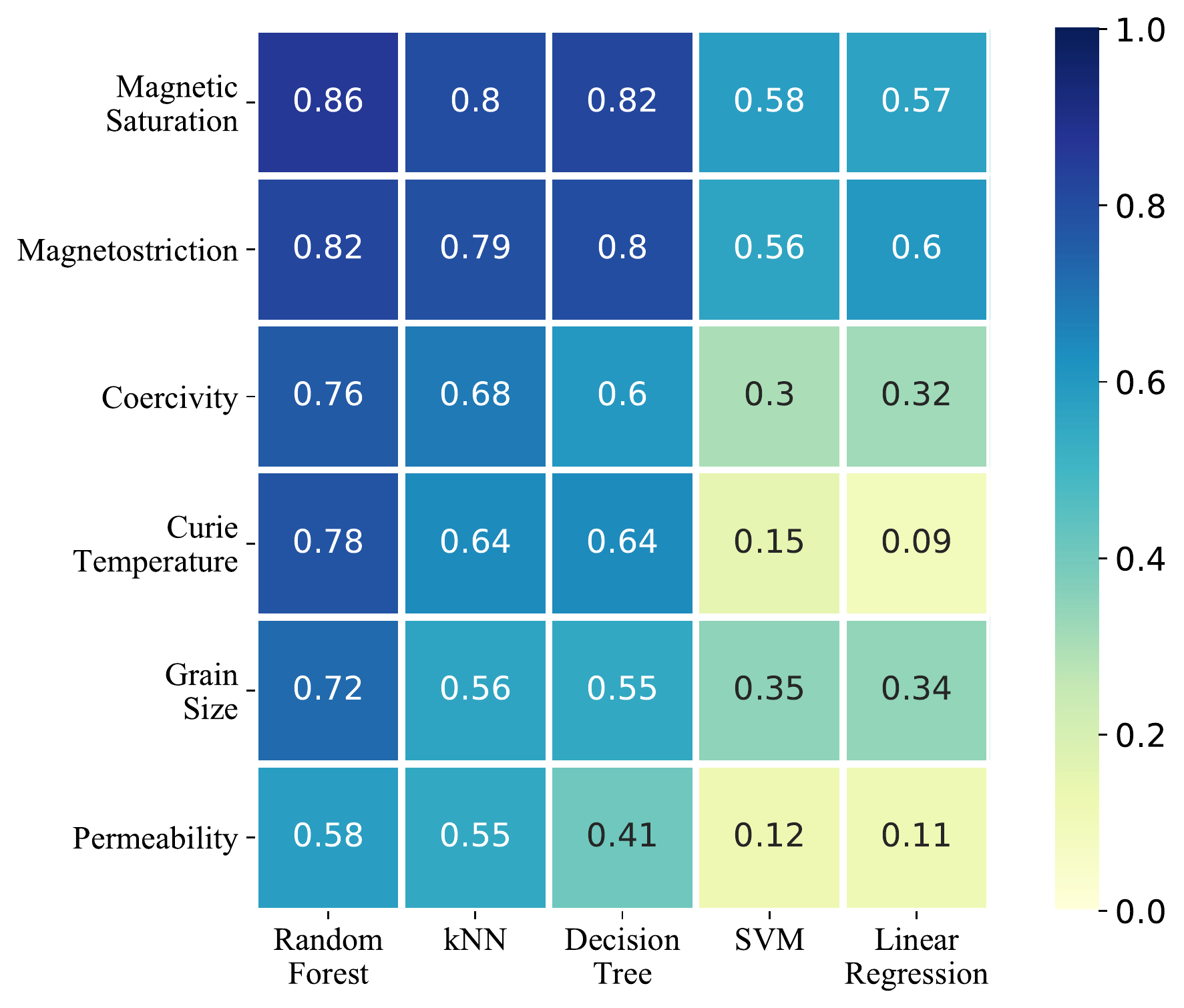}
    \caption{$R^2$ value of different machine learning algorithms. }
    \label{fig:heatmap}
\end{figure}

Five different machine learning algorithms including linear regression, support vector machines, decision trees, k-nearest neighbors and random forest, were utilized and compared with each other in the process of building predictive models \cite{pedregosa2011scikit}. The comparison of the coefficient of determination ($R^2$) score from 20-fold cross-validation is shown in Fig.~\ref{fig:heatmap}. 
Note that we performed a natural log transformation on coercivity to fix its skewness. Based on the $R^2$ score, it is evident that the random forest model is the best. Random forest regression(RFR) \cite{breiman2001random} is an ensemble of regression trees, induced from bootstrap samples of the training data, using random feature selection in the tree induction process. Prediction is then made by averaging the outputs of the ensemble of trees. Random forest generally exhibits a substantial performance improvement over the single tree classifier, such as CART and C4.5. It yields a generalization error rate that compares favorably to AdaBoost, yet is more robust to noise \cite{chen2004using}. 

\begin{figure}
    \centering
    \includegraphics[width=\textwidth]{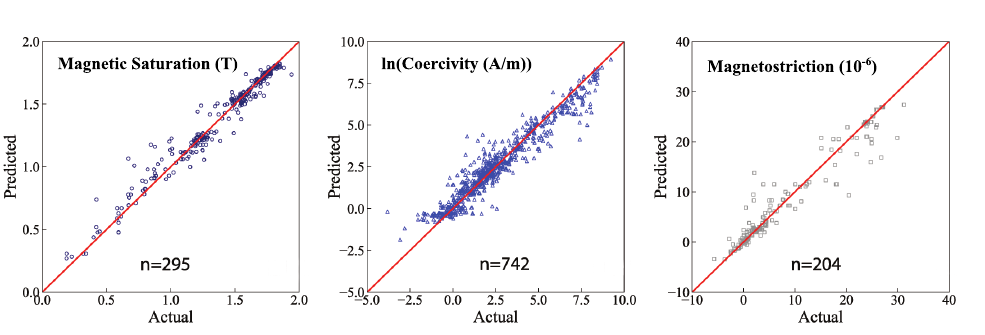}
    \caption{Comparison between predicted values from the machine learning model and experimental values for (a) magnetic saturation, (b) coercivity, and (c) magnetostriction.}
    \label{prediction}
\end{figure}

Predicted values of magnetic saturation, coercivity and magnetostriction from random forest models are compared with actual experimental values in Fig.~\ref{prediction}.
The resulting RFR achieves a good agreement with the experimentally measured properties. Investigating the regression data closer, it is apparent that the RFR model underestimates coercivity within the low coercivity region. This is potentially due to the high sensitivity of coercivity to the processing conditions. Furthermore, for all three properties, the data is not perfectly uniformly distributed across the entire range of interest. The discrepancy between predictions and measurements likely arises from a lack of data within the corresponding regions. 

\section{Material design and optimization}
Based on the predictive models built above, we attempted to find optimized material compositions and heat treatment conditions that can achieve large magnetic saturation and minimum loss in the presence of trade-offs between these two conflicting objectives. In the materials informatics field, machine learning models often have limitations when predicting beyond the limits of training data \cite{kauwe2020can,xiong2020evaluating}, and furthermore the random forest model is mathematically unable to extrapolate. Therefore, our main goal here is to achieve better balance between incompatible properties near the Pareto front and discover novel materials that can achieve similar or better performance to FINEMET. To achieve this objective, we performed two rounds of optimizations incrementally using the differential evolution algorithm. Experimental measurements conducted after the first round of optimization was added to the data set to try to improve the results in the second round. For the first round, the input space of the optimization process was the combination of features in the coercivity model, magnetostriction model, and magnetic saturation model, which are shown in Table \ref{table:feature selection}. For the second round, the input space of the optimization process was the combination of features in the coercivity model and magnetic saturation model.

\subsection{Differential Evolution}
Differential Evolution (DE) \cite{storn1997differential} is a stochastic population based method that is useful for global optimization problems. It utilizes NP D-dimensional parameter vectors
\begin{equation}
    x_{i,G}, i=1,2,...,\textnormal{NP}
\label{parameter vectors}
\end{equation}
as a population for each generation $G$. The initial vector population is chosen randomly, assuming a uniform probability distribution. DE generates new parameter vectors by adding the weighted difference between two population vectors to a third vector, which is called mutation. The mutated solution is mixed with other candidate solutions to create a trial candidate. In this study, the ``best1bin" strategy was utilized. In this strategy,  two members of the population are randomly chosen. Their difference is used to mutate the trial candidate
\begin{equation}
    \mu_{i,G+1} =x_{r_1,G}+ F\cdot(x_{r_2,G}-x_{r_3,G}),
\end{equation}
while $r_1, r_2, r_3 \in \{1,2,..., \textnormal{NP}\}$ are random different indexes and $F$ is a constant parameter $\in [0,2]$ that controls the amplitude of the differential variation $(x_{r_2,G}-x_{r_3,G})$. 
In order to increase the diversity of the perturbed parameter vectors, a crossover is introduced. To this end, the trial candidate
\begin{equation}
    \nu_{i,G+1}=(\nu_{1i,G+1}, \nu_{2i,G+1}, ..., \nu_{Di,G+1})
\end{equation}
is formed, where $\nu_{ji, G+1}$ $(j=1,2,...,D)$ is determined by a binomial distribution. A random number $\in [0, 1)$ is generated, if this number is less than the recombination constant that is determined by the user, then $\nu_{ji, G+1}$ is loaded from $\mu_{ji, G+1}$, otherwise it is loaded from the original candidate $x_{ji, G+1}$. It is ensured that $\nu_{i,G+1}$ gets at least one parameter from $\mu_{i, G+1}$.
The choice of whether to use trial candidate $\nu_{i,G+1}$ or the original candidate $x_{i, G+1}$ is made using the greedy criterion. Once the trial candidate is built, its fitness is assessed. If the trial is better than the original candidate, then it takes its place. If it is also better than the best overall candidate, it replaces that value too. 

\subsection{Optimization Results}
Our primary goal in designing improved Fe-based soft magnetic alloys is to minimize the core loss to help reduce the energy waste during operation. A secondary goal is to maximize the magnetic saturation. To ensure the existence of Fe-Si phase, a constraint was employed to ensure the atomic percentage of Si was no less than 3\%. Both design targets can contribute to the design of high energy efficiency. Properties like coercivity and magnetostriction, which directly affect the core loss could serve as our targets for loss minimization.

The first choice is to formulate the problem as a single-objective optimization with the objective function being the magnetic saturation. To satisfy the prerequisite of low core loss, we reformulated two objectives as constraints to restrict the design space while maximizing the objective function. The constraints were described as $\ln$(coercivity) not exceeding $C_0$ (A/m) and magnetostriction not exceeding $M_0$ ($\times10^{-6}$). 

The second choice is to formulate the problem as a multi-objective optimization. We only reformulate magnetostriction as the constraint and define a composite objective function to be minimized as:
\begin{equation}
    V=-\alpha_1 (B_S)+\alpha_2\ln(H_C) 
\end{equation}
where $\alpha_1$ and $\alpha_2$ values the importance of each of the properties to achieve a balance between conflicting objectives.  
The first round of optimization emphasized achieving a low coercivity and the second round emphasized achieving a high magnetic saturation. For the first round, the optimization ran on the composition space of Fe, Si, Cu Ta, Mo, Nb, Zr, B, and P. We constrained our composition space further so that Si was larger than 3\% to ensure the existence of the Fe$_3$Si phase. We tried a total of four different strategies of single-objective and multi-objective approaches. In single-objective methods, the value of $C_0$ was chosen to be $-1.5$ or $-0.5$ and $M_0$ was chosen to be 3. In multi-objective methods, $M_0$ was also set to 3 and two different weight combinations have been explored in our calculations: $\alpha_1=\alpha_2=1$ and $\alpha_1=4$, $\alpha_2=1$. 
For the second round, we added the additional experimental results, which we generated following the first round of optimization to our database and re-trained the machine learning model. The second round of optimization ran on a smaller composition space of Fe, Si, Cu, Mo, Nb, B, Ge, and P where Si needed to be larger than 3\%. In this round, our model only ran on coercivity and magnetic saturation because of the high correlation between coercivity and magnetostriction. For the second round, we tried the single-objective approach and focused on maximizing magnetic saturation with constraints of $C_0$ to be two separate values of 0.5 or 0.  The problem formulations of both first and second round optimizations are shown in \Cref{table:first round problem multi,table:first round problem single,table:second round problem}.

\begin{table}[]
\centering
\begin{tabular}{ll}
\hline\hline
\multicolumn{1}{l|}{Feature space} & Fe, Si, Cu, Ta, Mo, Nb, Zr, B, P, $T_a$, $t_s$, ln($H_C$), $\lambda$, $B_S$ \\ \hline
\multicolumn{1}{l|}{Objective function} &  $V=-\alpha_1 (B_S)+\alpha_2\ln(H_C)$ \\
\hline
\multicolumn{1}{l|}{Constraints} &  Si $>$ 3\%; $\lambda$ $<$ 3 ($\times10^{-6}$); \\
\multicolumn{1}{l|}{} & Ta, Mo, Nb, Zr could be constrained to zero in certain cases.\\
\hline\hline     
\end{tabular}
\caption{First round multi-objective strategy problem formulation. Where $T_a$ is annealing temperature (K) and $t_s$ is annealing time (s).}
\label{table:first round problem multi}
\end{table}

\begin{table}[]
\centering
\begin{tabular}{ll}
\hline\hline
\multicolumn{1}{l|}{Feature space} & Fe, Si, Cu, Ta, Mo, Nb, Zr, B, P, $T_a$, $t_s$, ln($H_C$), $\lambda$, $B_S$ \\ \hline
\multicolumn{1}{l|}{Objective function} &  $V=-B_S$ \\
\hline
\multicolumn{1}{l|}{Constraints} &  Si $>$ 3\%; $\lambda$ $<$ 3 ($\times10^{-6}$); ln(Hc) $<$ -1.5 or 0.5; \\
\multicolumn{1}{l|}{} & Ta, Mo, Nb, Zr could be constrained to zero in certain cases.\\
\hline\hline     
\end{tabular}
\caption{First round single-objective strategy problem formulation. }
\label{table:first round problem single}
\end{table}

\begin{table}[]
\centering
\begin{tabular}{ll}
\hline\hline
\multicolumn{1}{l|}{Feature space} & Fe, Si, Cu, Mo, Nb, B, Ge, P, $T_a$, $t_s$, ln$(H_C)$, $B_S$ \\ \hline
\multicolumn{1}{l|}{Objective function} &  $V=-B_S$ \\
\hline
\multicolumn{1}{l|}{Constraints} &  Si $>$ 3\%; Fe $>$ 75\%; ln(Hc) $<$ 0 or 0.5; \\
\multicolumn{1}{l|}{} & Mo, Ge, P could be constrained to zero in certain cases.\\
\hline\hline     
\end{tabular}
\caption{Second round problem formulation. }
\label{table:second round problem}
\end{table}

\begin{figure}
    \centering
    \includegraphics[width=\textwidth]{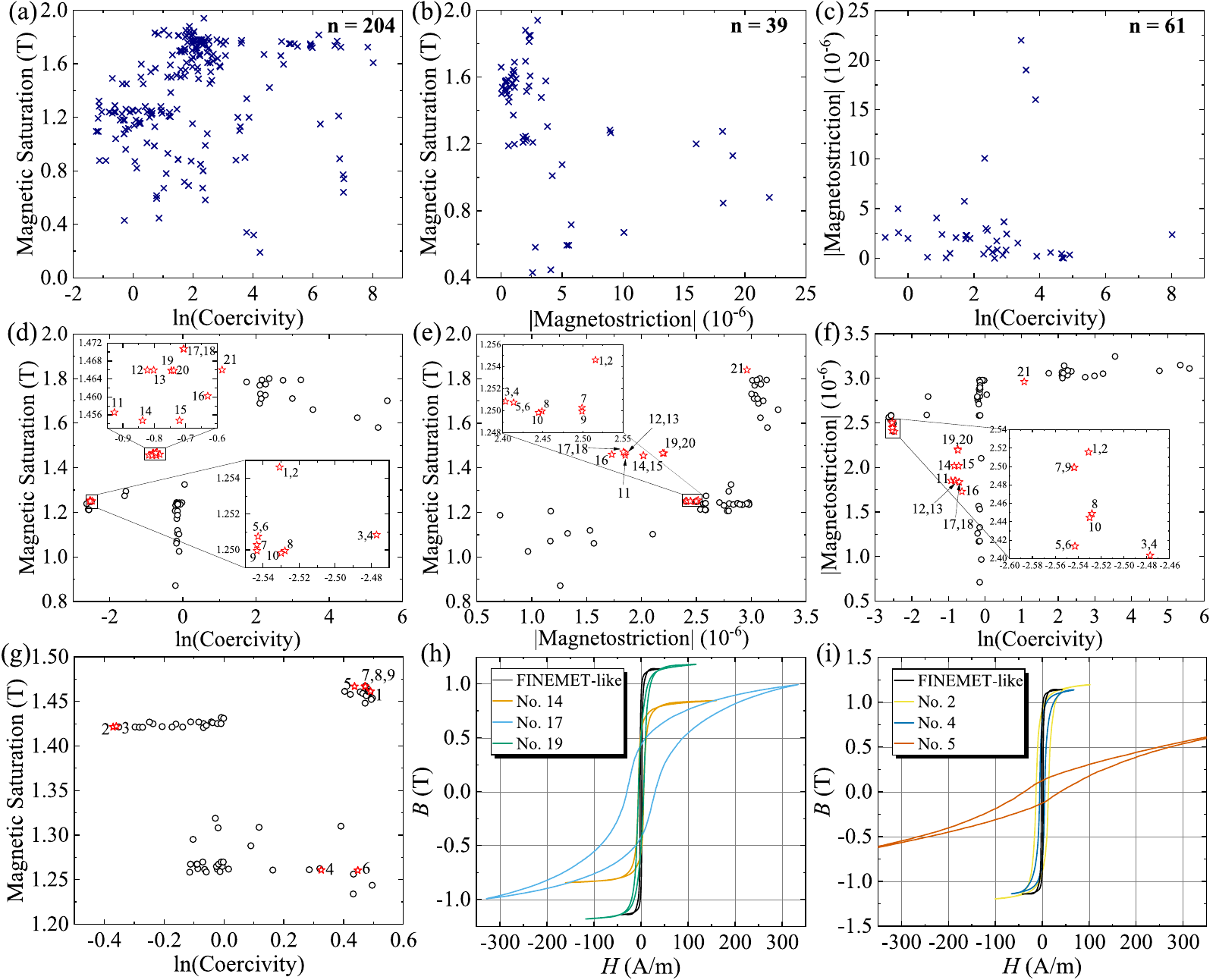}
    \caption{Trade-off surface of different combinations of magnetic properties. (a)-(c) are from the experimental data set and (d)-(g) are from the machine learning optimizations. (d)-(f) are for first iteration and (g) is for second iteration.  $n$ is the number of points contained in the plots. The selected results marked by index are shown in Table \ref{table:results}. (h) and (i) are experimental measurements of $B$-$H$ loops for different samples and compared with a commercial FINEMET-like alloy composition processed similar to the experimental alloys investigated. (h) is for first iteration and (i) is for second iteration.}
    \label{design}
\end{figure}

Fig.~\ref{design}
shows the different combination of trade-off surface plots for the results of all the optimization methods we tried, including both single-objective and multi-objective methods. Fig.~\ref{design}(a)-(c) are the plots of experimental measurements from the database and Fig.~\ref{design}(d)-(g) represent the optimization results achieved by applying DE optimization to the RFR models. When there are two or more objectives, solutions rarely exist that optimize all at once. The objectives are normally measured in different units, and any improvement in one is at the loss of another \cite{ashby2000multi}. It can be seen in Fig.~\ref{design}(d)-(g) that there is a systematic trade-off between coercivity and magnetostriction, versus magnetic saturation. Increasing coercivity and magnetostriction generally led to an increase of magnetic saturation. 

Solutions marked by numbers and shown in Table \ref{table:results} and Table \ref{table:results2}
\begin{table*}
\centering
{\scriptsize
\begin{tabular}{>{\centering}p{0.42cm}>{\centering}p{0.42cm}>{\centering}p{0.42cm}>{\centering}p{0.42cm}>{\centering}p{0.42cm}>{\centering}p{0.42cm}>{\centering}p{0.42cm}>{\centering}p{0.42cm}>{\centering}p{0.42cm}>{\centering}p{0.42cm}>{\centering}p{0.42cm}>{\centering}p{0.42cm}>{\centering}p{0.7cm}>{\centering}p{0.5cm}>{\centering}p{0.42cm}>{\centering}p{0.42cm}>{\centering}p{0.42cm}>{\centering}p{0.7cm}}
\hline\hline
Index & Fe & Si & Cu & Ta & Mo & Nb & Zr & B  & P & $T_a$ & $t_s$ & ln($H_C$) & $\lambda$  & $B_S$ & $\alpha_1$ & $\alpha_2$ & $C_0$ \tabularnewline
\hline
1 & 74.09 & 13.67 & 0.56 & 0 & 0 & 3.05 & 0.36 & 8.14 & 0.14 & 817 & 3959 & $-$2.53 & 2.52 & 1.25 & 4 & 1 & Inf\tabularnewline
2 & 74.12 & 13.38 & 0.71 & 0 & 0 & 2.69 & 0.76 & 8.25 & 0.09 & 816 & 4660 & $-$2.53 & 2.52 & 1.25 & 4 & 1 & Inf \tabularnewline
3 & 73.86 & 14.22 & 0.66 & 0 & 0 & 2.69 & 0 & 8.19 & 0.37 & 816 & 1206 & $-$2.48 & 2.40 & 1.25 & 4 & 1 & Inf\tabularnewline
4 & 73.93 & 14.22 & 0.62 & 0 & 0 & 2.79 & 0 & 8.14 & 0.30 & 816 & 121 & $-$2.48 & 2.40 & 1.25 & 4 & 1 & Inf \tabularnewline
5 & 73.74 & 14.18 & 0.52 & 0 & 0 & 2.95 & 0 & 8.14 & 0.48 & 817 & 4285 & $-$2.54 & 2.41 & 1.25 & 1 & 1 & Inf  \tabularnewline
6 & 73.96 & 14.16 & 0.61 & 0 & 0 & 2.61 & 0 & 8.31 & 0.35 & 817 & 3609 & $-$2.54 & 2.41 & 1.25 & 1 & 1 & Inf  \tabularnewline
7 & 73.53 & 13.35 & 0.68 & 0 & 0.17 & 3.43 & 0.24 & 8.51 & 0.09 & 816 & 5052 & $-$2.54 & 2.50 & 1.25 & 1 & 1 & Inf \tabularnewline
8 & 73.68 & 13.30 & 0.70 & 1.09 & 0.05 & 2.78 & 0.02 & 8.23 & 0.15 & 816 & 6086 & $-$2.53 & 2.45 & 1.25 & 1 & 1 & Inf \tabularnewline
9 & 74.01 & 13.47 & 0.65 & 0.50 & 0 & 2.75 & 0.20 & 8.30 & 0.11 & 817 & 3490 & $-$2.54 & 2.50 & 1.25 & 1 & 1 & Inf \tabularnewline
10 & 73.40 & 14.21 & 0.69 & 0.35 & 0.36 & 2.72 & 0 & 8.02 & 0.26 & 816 & 5091 & $-$2.53 & 2.44 & 1.25 & 1 & 1 & Inf \tabularnewline
11 & 77.13 & 9.04 & 0.29 & 0.97 & 0.01 & 3.29 & 1.17 & 7.92 & 0.19 & 783 & 907 & $-0.93$ & 1.85 & 1.46 & 1 & 0 & -0.5  \tabularnewline
12 & 78.04 & 9.19 & 0.31 & 0.59 & 0 & 3.45 & 0 & 8.15 & 0.28 & 784 & 4290 & $-0.82$ & 1.85 & 1.47 & 1 & 0 & -0.5 \tabularnewline
13 & 79.08 & 9.26 & 0.29 & 0 & 0.21 & 3.33 & 0 & 7.80 & 0.04 & 784 & 3196 & $-0.80$ & 1.85 & 1.47 & 1 & 0 & -0.5 \tabularnewline
14 & 76.67 & 8.45 & 0.37 & 0.58 & 0 & 3.12 & 2.81 & 7.59 & 0.42 & 786 & 3489 & $-0.84$ & 2.02 & 1.45 & 1 & 0 & -0.5 \tabularnewline
15 & 76.62 & 8.35 & 0.59 & 0 & 1.44 & 2.55 & 2.41 & 7.92 & 0.12 & 784 & 4356 & $-0.72$ & 2.02 & 1.45 & 1 & 0 & -0.5 \tabularnewline
16 & 78.49 & 8.53 & 0.47 & 1.56 & 0.28 & 2.57 & 0 & 8.04 & 0.06 & 783 & 991 & $-0.63$ & 1.73 & 1.46 & 1 & 0 & -0.5 \tabularnewline
17 & 79.90 & 8.96 & 0.61 & 0 & 0 & 2.54 & 0 & 7.98 & 0.02 & 785 & 1145 & $-$0.71 & 1.84 & 1.47 & 1 & 0 & -0.5 \tabularnewline
18 & 79.79 & 8.92 & 0.57 & 0 & 0 & 2.47 & 0 & 7.99 & 0.26 & 784 & 1066 & $-$0.71 & 1.84 & 1.47 & 1 & 0 & -0.5 \tabularnewline
19 & 78.69 & 8.80 & 0.54 & 0 & 0 & 2.50 & 1.32 & 7.92 & 0.23 & 784 & 6038 & $-$0.75 & 2.21 & 1.47 & 1 & 0 & -0.5 \tabularnewline
20 & 79.16 & 8.87 & 0.63 & 0 & 0 & 2.57 & 0.52 & 7.92 & 0.34 & 783 & 5436 & $-$0.74 & 2.19 & 1.47 & 1 & 0 & -0.5 \tabularnewline
21 & 83.08 & 4.23 & 1.11 & 0.91 & 0 & 0 & 0 & 7.09 & 3.58 & 723 & 1159 & 1.07 & 2.96 & 1.84 & 4 & 1 & Inf  \tabularnewline
\hline\hline
\end{tabular}}
\caption{Selected first round Optimization results obtained by DE using Random Forest model. Where $T_a$ is annealing temperature (K) and $t_s$ is annealing time (s). $\alpha_1$, $\alpha_2$, $C_0$ are optimization parameters, Inf means there's no constraint on coercivity. Constraint on magnetostriction($M_0$) is always set to be 3.}
\label{table:results}
\end{table*}
are identified as optimum solutions based on the trade-off surface described by the $\ln$(coercivity)-magnetic saturation plot, as these two are defined as our main objectives. Further selection in the optimum set could be based on different application scenarios and different weighting strategies of the two competing aspects.
\begin{table*}
\centering
{\scriptsize
\begin{tabular}{>{\centering}p{0.7cm}>{\centering}p{0.55cm}>{\centering}p{0.55cm}>{\centering}p{0.55cm}>{\centering}p{0.55cm}>{\centering}p{0.55cm}>{\centering}p{0.55cm}>{\centering}p{0.55cm}>{\centering}p{0.55cm}>{\centering}p{0.55cm}>{\centering}p{0.55cm}>{\centering}p{0.7cm}>{\centering}p{0.7cm}>{\centering}p{0.7cm}>{\centering}p{0.7cm}}
\hline\hline
Index & Fe & Si & Cu& Mo& Nb& B & Ge& P & $T_a$ & $t_s$ & ln$(H_C)$ & $B_S$ & $C_0$\tabularnewline
\hline
1 & 76.23 & 11.95 & 0.30 & 0.00 & 2.25 & 8.83 & 0.41 & 0.04 & 785 & 2862 & -0.37 & 1.42  & 0 \tabularnewline
2 & 76.31 & 11.96 & 0.33 & 0.21 & 2.21 & 8.99 & 0.00 & 0.00 & 783 & 2585 & -0.36 & 1.42  & 0 \tabularnewline
3 & 76.66 & 11.88 & 0.46 & 0.00 & 2.25 & 8.70 & 0.00 & 0.04 & 783 & 2938 & -0.35 & 1.42  & 0 \tabularnewline
4 & 76.97 & 11.51 & 0.45 & 0.00 & 2.29 & 8.78 & 0.00 & 0.00 & 663 & 3150 & 0.44  & 1.47  & 0.5 \tabularnewline
5 & 77.04 & 11.29 & 0.35 & 0.00 & 2.35 & 8.86 & 0.12 & 0.00 & 661 & 3078 & 0.44  & 1.47  & 0.5 \tabularnewline
6 & 76.62 & 11.43 & 0.49 & 0.00 & 2.43 & 8.58 & 0.44 & 0.01 & 662 & 3021 & 0.47  & 1.47  & 0.5 \tabularnewline
7 & 76.67 & 11.67 & 0.38 & 0.18 & 2.26 & 8.56 & 0.28 & 0.00 & 663 & 2766 & 0.47  & 1.47  & 0.5 \tabularnewline
8 & 76.88 & 11.43 & 0.42 & 0.14 & 2.34 & 8.80 & 0.00 & 0.00 & 662 & 3132 & 0.47  & 1.47  & 0.5 \tabularnewline
9 & 76.74 & 11.87 & 0.48 & 0.02 & 2.29 & 8.61 & 0.00 & 0.00 & 663 & 3201 & 0.47  & 1.47  & 0.5\tabularnewline                                                         
\hline\hline           
\end{tabular}}
\caption{Selected second round optimization results obtained by DE using Random Forest model.}
\label{table:results2}
\end{table*}

\begin{figure}
    \centering
    \includegraphics[scale=0.8]{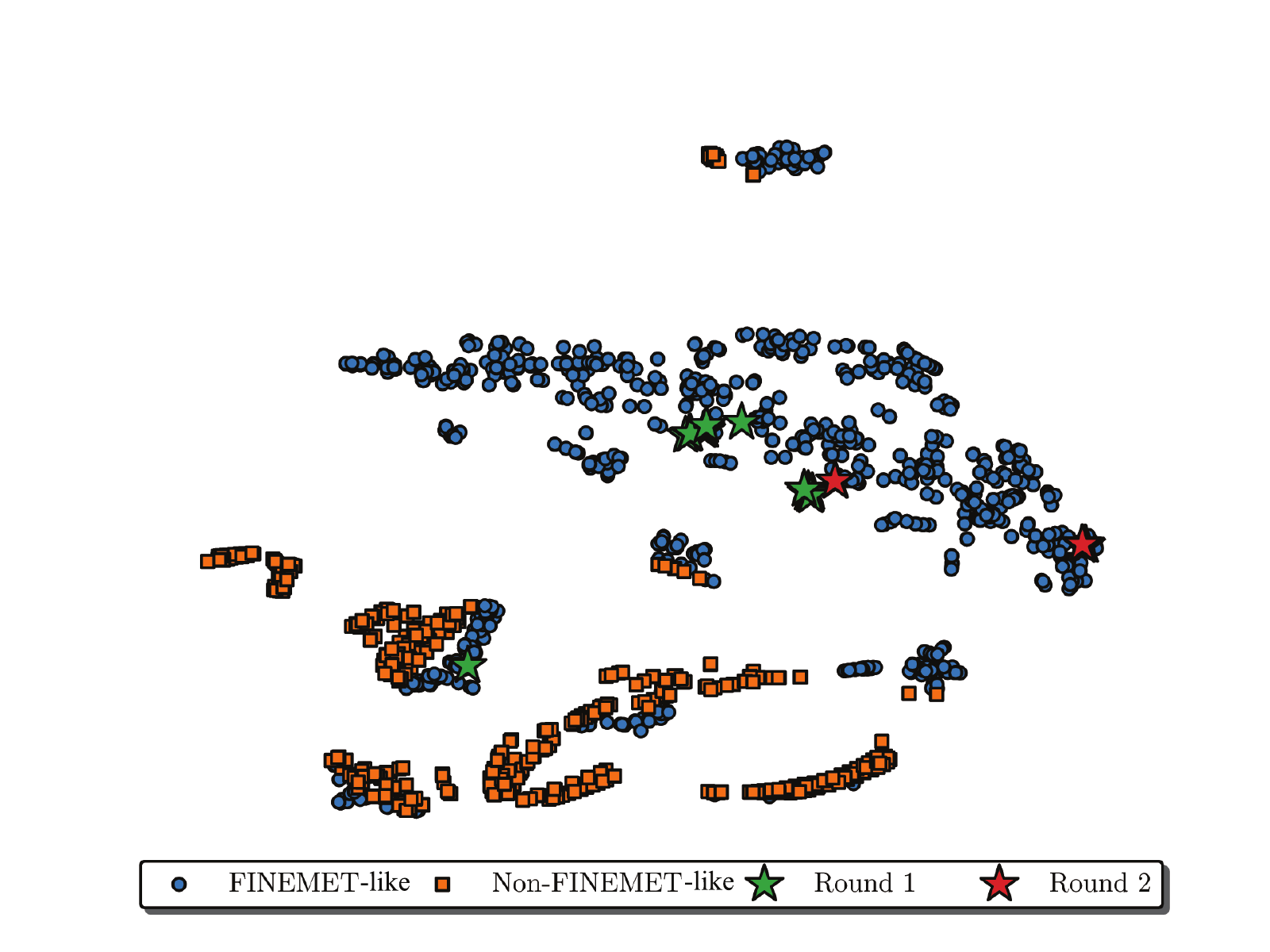}
    \caption{A t-SNE mapping of the compositions and processing conditions of the alloys in the literature database as well as the alloys found from the first and second rounds of optimization. Note that the precise positions of alloys in this plot are not quantitatively meaningful.}
    \label{tsne}
\end{figure}

Two-dimensional t-distributed stochastic neighbor embedding (t-SNE) was used to visualize how the optimized alloys compare to the ones presented in the published literature. Fig.~\ref{tsne} displays this t-SNE mapping, which was conducted on the processing space of the literature database as well as the alloys found in the first and second rounds of optimization, shown in Tables \ref{table:results} and \ref{table:results2} respectively. In this plot, the processing space was defined as the composition of every element in each alloy as well as the annealing temperature and time, scaled by their respective maximum and minimum values in the database. For the purpose of visualization, ``FINEMET-like'' alloys were defined as alloys containing greater than 3\% Si while ``Non-FINEMET-like'' alloys contain less Si. These groups seem rather separable in Fig.~\ref{tsne}, indicating that they represent two distinct classes of materials in the literature. The alloys found in both rounds of optimization are shown among the FINEMET-like alloys, which is to be expected given the constraints on Si and Fe shown in Tables~\ref{table:first round problem multi}, \ref{table:first round problem single}, and \ref{table:second round problem}. While the optimized alloys are not shown to exist entirely separate from literature data, they do not match any alloy from literature exactly. Optimized alloys also appear to lie near the bounds of their respective groups, indicating that further iterations of optimal experimental design could expand the boundaries of current knowledge in this space.

\section{Experimental validation}

To experimentally validate our machine learning model, several predicted compositions near the Pareto front of the ln(coercivity) - magnetic saturation plots have been synthesized. For the first round, we chose three points No.14, No.17 and No.19 in the points region of intermediate value of both coercivity and magnetic saturation as shown in Fig.~\ref{design}(d). While for the second round, we chose one point from each of the three point regions, namely No.2, No.4 and No.5, to try to test the behavior of different regions of point segregation in Fig.~\ref{design}(g). 

The alloys produced for this study were melted from elemental constituents and flipped and remelted several times to ensure homogeneity and then cast into cigar-shaped ingots. The ingots weighed approximately 60 g. The ingots were then used as melt stock for melt spinning using an Edward Buehler HV melt spinner using a planar flow casting process and wheel speed of 25.9 m/s. The cast ribbons were approximately 19.5 microns thick and $\sim$16.5 mm wide. Composition of all melt-spun ribbons were confirmed by inductively coupled plasma atomic emission spectroscopy (ICP-AES). The ribbon was wound into small cores, wrapped with a piece of copper wire to hold the core together and then heat-treated in argon after first pulling a  vacuum to approximately $1\times10^{-7}$ torr and heat-treated at the times/temperatures specified. The heating rate was ~3 $^{\circ}$C/min and samples were cooled at a rate of 8 $^{\circ}$C/min after the specified treatment. Properties of the wound cores were determined by IEEE Standard 393-1991: Standard for Test Procedures for Magnetic Materials. Testing was performed at 1000 Hz. 

From the resulting $B$-$H$ loops as shown in Figs.~\ref{design}(h) and (i), magnetic properties were determined, including $B_S$ and $H_C$. The comparison of predicted and experimental properties of selected samples from both the first iteration and second iteration are shown in Table \ref{table:validation}. The compositions are recorded by experimental measurement of the prepared samples and the heat treatment times and temperatures are the same as the model predictions and listed in Tables \ref{table:results} and \ref{table:results2}. It should be noted from Figs.~\ref{design}(h) and (i) that sample No.19 in first round and No.2 in second round have similar performance compared to commercial FINEMET-like alloys, which shows that our approach is effective in identifying other compositions with very good properties.

We collected a substantial portion of the reported experimental data, which is shown in Fig.~\ref{design}(a)-(c) as Ashby plots. Although there are a large number of data points describing ln(coercivity)-magnetic saturation property space  in Fig.~\ref{design}(a), it can still be observed that the high-$B_S$, low-$H_C$ area (top left corner) is completely empty. The ultimate goal of the material design is to breach the boundary and reach the target area. To improve the machine learning model, we need more data, and an efficient way is to explore the property space that is missing, if at all possible. Another potential strategy is to continue to iterate the design process. By using experimental results of proposed compositions associated with better properties it may be possible to continue to iterate until the target properties are obtained. 

As a complex material system, soft magnetic nanocrystalline alloys usually contain several elements, which makes it difficult to decide what elements should be included. Several principles could be considered: (a) For applications of soft magnetic materials, the prices of elements are non-negligible factors. In our design process, the selection of late transition metal is a typical issue due to the hefty price difference between Au and Cu. The price of Cu is about 6 USD/kg, Ag about 500 USD/kg and Au over 40,000 USD/kg. (b) If possible, we try not to include too many different elements that serve the same function in the alloy. (c) In this work, the only magnetic element included is Fe, which can provide a pure $\alpha$-Fe$_3$Si nanocrystalline phase. Based on these principles, we proposed the optimized compositions listed in Table~\ref{table:results} and Table~\ref{table:results2} for potential experimental validations, in addition to the selected alloys confirmed in Table~\ref{table:validation}. Note that, as shown in Table~\ref{table:validation}, there are discrepancies between predicted and measured values, which probably arises due to several facts: (a) We have been predicting properties in a high-dimensional space without adequate data; (b) Random forest models have limitations for predicting extreme values. Modern development in machine learning algorithms such as attention-based network \cite{Wang2020} could be employed to improve the performance. By running more iterations of our optimized experimental design framework, the models will become more robust in predicting outstanding magnetic properties.

\begin{table*}
\centering
{\scriptsize
\begin{tabular}{>{\centering}p{1cm}>{\centering}p{5.5cm}>{\centering}p{1.5cm}>{\centering}p{1.5cm}>{\centering}p{1.5cm}>{\centering}p{1.5cm}}
\hline\hline
Index & Composition & \multicolumn{2}{c}{ln$(H_C)$ (A/m)} & \multicolumn{2}{c}{$B_S$ (T)} \tabularnewline
\hline
\multicolumn{6}{c}{Round 1}\tabularnewline
 & & \underline{Predicted} & \underline{Measured} & \underline{Predicted} & \underline{Measured} \tabularnewline
 14 & Fe$_{77.1}$Si$_{8.5}$Cu$_{0.4}$Nb$_{3.1}$Ta$_{0.6}$B$_{7.3}$P$_{0.3}$Zr$_{2.8}$ & -0.82 & 2.19 & 1.455 & 0.915 \tabularnewline
 17 & Fe$_{80}$Si$_{9}$Cu$_{0.7}$Nb$_{2.5}$B$_{7.9}$ & -0.69 & 3.39 & 1.471 & 0.913 \tabularnewline
 19 & Fe$_{78.9}$Si$_{8.9}$Cu$_{0.6}$Nb$_{2.5}$B$_{7.7}$P$_{0.2}$Zr$_{1.2}$ & -0.76 & 1.79 & 1.466 &1.146 \tabularnewline
\hline
\multicolumn{6}{c}{Round 2}\tabularnewline
 & & \underline{Predicted} & \underline{Measured} & \underline{Predicted} & \underline{Measured} \tabularnewline
 2 & Fe$_{76.3}$Si$_{12}$Cu$_{0.3}$Nb$_{3.1}$Mo$_{0.2}$Nb$_{2.2}$B$_{9}$ & -0.36 & 2.58 & 1.42 & 1.195 \tabularnewline
 4 & Fe$_{77}$Si$_{11.5}$Cu$_{0.4}$Nb$_{2.3}$B$_{8.8}$ & 0.44 & 3.25 & 1.47 & 1.088 \tabularnewline
 5 & Fe$_{77}$Si$_{11.3}$Cu$_{0.4}$Nb$_{2.4}$B$_{8.8}$Ge$_{0.1}$ & 0.44 & 3.64 & 1.47 &1.014 \tabularnewline
\hline\hline
\end{tabular}}
\caption{Predicted and measured properties of soft magnetic alloys from the first and second round of optimization. The index corresponds to the value in Tables \ref{table:results} and \ref{table:results2}. Compositions are the experimental values from samples. Heat treatment time and temperature are referenced from Tables \ref{table:results} and \ref{table:results2}.}
\label{table:validation}
\end{table*}

\section{Summary}

In this work, we built a general database for Fe-based FINEMET-type soft magnetic nanocrystalline alloys using experimental data from all available literature. Based on this, machine learning techniques were applied to analyze the statistical inference of different features and then build predictive models to establish the relation between materials properties and material compositions along with processing conditions. We chose the random forest model as our modeling tool due to its better performance compared with several other machine learning methods. Optimization process has been performed to establish and then solve the inverse problem that is to find a suitable combination of element components and processing conditions to achieve minimum loss and maximum magnetic saturation. Experimental validations have been applied on several predicted materials, which showed that the predicted novel material could have similar performance as the commercial FINEMET-like alloys. Furthermore, the collected data set and analysis procedure can create more insight on how to design the next-generation optimized Fe-based soft magnetic nanocrystalline alloys motivated by various applications. The data set and analysis code are available on Github \cite{github}.  

\section*{Acknowledgments}

The authors would like to acknowledge the Data-Enabled Discovery and Design of Energy Materials (D$^3$EM) program funded through the National Science Foundation (NSF) Award No. 1545403. Y.W. also acknowledges the financial support of the NSF through Award No. 1508634. and 1905325. Y.T. also acknowledges the financial support of the College of Science Strategic Transformative Research Program (COS-STRP) at Texas A\&M University and the Robert A. Welch Foundation, Grant No. A-1526. O.L. also acknowledges the support of the NSF REU program through Award No. 1461202. R.D.N and V.K. acknowledge support from the Electrified Aircraft Powertrain Technologies Subproject in the NASA Advanced Air Vehicles Program, Amy Jankovsky subproject manager.

\bibliographystyle{elsarticle-num}

\bibliography{soft_magnetic}

\end{document}